\documentstyle[preprint,eqsecnum,aps]{revtex}
\tightenlines                                                                                                                                                                
\newcommand{\be}{\begin{equation}}
\newcommand{\ee}{\end{equation}}
\newcommand{\bea}{\begin{eqnarray}}
\newcommand{\eea}{\end{eqnarray}}
\newcommand{\ba}{\begin{array}}
\newcommand{\ea}{\end{array}}

\newcommand{\Th}{\Theta}
\newcommand{\th}{\theta}

\newcommand{\de}{\delta}

\newcommand{\pa}{\partial}

\newcommand{\no}{\nonumber}

\newcommand{\str}{\mbox{str}}

\begin{document}

\title{A Note on the Gauge Equivalence between the Manin-Radul \\
and Laberge-Mathieu Super KdV Hierarchies}

\author{Jiin-Chang Shaw$^1$ and Ming-Hsien Tu$^2$ }
\address{
$^1$ Department of Applied Mathematics, National Chiao Tung University, \\
Hsinchu, Taiwan, \\
and\\
$^2$ Department of Physics, National Tsing Hua University, \\
Hsinchu, Taiwan
}
\date{\today}
\maketitle

\begin{abstract}
The gauge equivalence between the Manin-Radul and Laberge-Mathieu
super KdV hierarchies is revisited. Apart from the Inami-Kanno transformation,
we show that there is another gauge transformation which also possess the 
canonical property. We explore the relationship of these two gauge
transformations from the Kupershmidt-Wilson theorem viewpoint
and, as a by-product, obtain the Darboux-Backlund transformation
for the Manin-Radul super KdV hierarchy.  The geometrical intepretation 
of these transformations is also briefly discussed.
\end{abstract}

\newpage

\section{Introduction}

Recently, Morosi and Pizzocchero \cite{MP1,MP2,MP3} discussed the gauge
equivalence of the Manin-Radul (MR) \cite{MR} and Laberge-Mathieu (LM) 
\cite{LM} super Korteweg-de Vries (sKdV) hierarchies from a bi-Hamiltonian
 and Lie superalgebraic viewpoint. This approach can be viewed as a 
superextension of the Drinfeld-Sokolov method \cite{DS} for building a
KdV-type hierarchy for a simple Lie algebra.
They showed \cite{MP1} that the gauge transformation proposed by 
Inami and Kanno (IK) \cite{IK} not only preserves the Lax formulations but also 
the bi-Hamiltonian structures corresponding to the MR and LM hierarchies. 
In particular, they provided an geometrical meaning of the IK transformation
which rests on the natural fibered structure appearing in
the bi-Hamiltonian reduction of loop superalgebras.

In this paper, in addition to the IK transformation, we find that
there is another gauge transformation between the MR and 
LM sKdV hierarchies preserving the Lax formulations.
We investigate the canonical property of this gauge transformation and 
discuss the connection to the IK transformation from the
Kupershmidt-Wilson theorem\cite{KW} viewpoint.  As a by-product,
the Darboux-B\"acklund transformation (DBT) for the MR sKdV
hierarchy can be constructed from these two gauge transformations.
The geometrical interpretation of these two transformations is also
briefly discussed.

Our paper is organized as follows: In Sec. II, the bi-Hamiltonian structures 
of the MR and LM sKdV hierarchies are briefly reviewed.
In Sec. III,  we introduce a gauge transformation between these two
hierarchies and investigate its canonical property. Then in Sec. IV,
we discuss the relationship between this transformation and
the IK transformation from the KW theorem viewpoint.   
Concluding remarks are presented in Sec. V. 

\section{Bi-Hamiltonian structures of the MR and LM sKdV hierarchies}

The MR sKdV hierarchy was defined originally from the reduction of the MR super 
Kadomtsev-Petviashvili hierarchy \cite{MR}. It has Lax equation as follows,
\be
\pa_nL^{MR}=[B^{MR}_n,L^{MR}],
\label{eqmr}
\ee
with
\bea
L^{MR}&=&\pa^2-\phi D+a
\label{laxmr}\\
B^{MR}_n&=&(-1)^n4^n(L^{MR})^{n+1/2}_+,
\eea
where the superderivative $D\equiv\pa_{\th}+\th\pa$ satisfies
$D^2=\pa$, $\th$ is the Grassmann variable ($\th^2=0$) which, together with
the even variable $x\equiv t_1$, define 
the $(1|1)$ superspace \cite{D} with coordinate $(x,\th)$.
The formal inverse of $D$ is introduced by $D^{-1}=\th+\pa_{\th}\pa^{-1}$,
which satisfies $DD^{-1}=D^{-1}D=1$. The coefficients 
$\phi=\phi(x,\th)$ and $a=a(x,\th)$ are an odd and an even superfield on
$(1|1)$ superspace, respectively. We denote the action of the superderivative $D$ on
the superfield $f$ by $(Df)$.

The bi-Hamiltonian structure of the MR hierarchy has been obtained in \cite{OP}as
\bea
\Th^{MR}_1& : &
\left(
\ba{c}
\de a\\
\de \phi
\ea
\right)
\rightarrow
\left(
\ba{c}
\dot{a}\\
\dot{\phi}
\ea
\right)=
\left(
\ba{cc}
-D\pa+\phi & -\pa\\
-\pa & 0
\ea
\right)
\left(
\ba{c}
\de a\\
\de \phi
\ea
\right)
\label{hmr1}\\
\Th^{MR}_2& : &
\left(
\ba{c}
\de a\\
\de \phi
\ea
\right)
\rightarrow
\left(
\ba{c}
\dot{a}\\
\dot{\phi}
\ea
\right)=
\left(
\ba{cc}
P_{aa} & P_{a\phi}\\
P_{\phi a} & P_{\phi\phi}
\ea
\right)
\left(
\ba{c}
\de a\\
\de \phi
\ea
\right),
\label{hmr2}
\eea
where the operators $P_{ij}$ are given by 
\bea
P_{aa}&=&D\pa^3-3\phi\pa^2+4aD\pa+(2(Da)-3\phi_x)\pa+2a_xD+3\phi(D\phi)\no\\
& &+(Da)_x-4a\phi-\phi_{xx}+\phi D^{-1}(Da)-(Da)D^{-1}\no\phi\\
& &-\phi D^{-1}\phi D^{-1}\phi-\phi D^{-1}\phi_x+\phi_x D^{-1}\phi\\
P_{a\phi}&=&
\pa^3-2\phi D\pa+4a\pa-\phi_xD+2a_x+\phi D^{-1}(D\phi)\\
P_{\phi a}&=&
\pa^3+2\phi D\pa+(4a-2(D\phi))\pa+\phi_x+2a_x-(D\phi)_x+(D\phi)D^{-1}\phi\\
P_{\phi\phi}&=&
4\phi\pa+2\phi_x.
\eea
Here, following the notations in \cite{MP1}, the phase space for the MR
theory is a pair $m=(a, \phi)$. A tangent vector at a point $m$ is denoted by
$\dot{m}=(\dot{a}, \dot{\phi})$ and a cotangent vector as a pair
$\de m=(\de a, \de \phi)$ where $\dot{a}$ and $\de \phi$ are even superfields, 
whereas $\dot{\phi}$ and $\de a$ are odd. The inner product is defined by
$\langle \de m, \dot{m}\rangle\equiv \int dx d\th (\de a\dot{a}+\de \phi\dot{\phi})$.

For LM hierarchy, the Lax equation is given by
\be
\pa_nL^{LM}=[B^{LM}_n,L^{LM}],
\label{eqlm}
\ee
with
\bea
L^{LM}&=&\pa^2-2u\pa-((Du)+\tau)D
\label{laxlm}\\
B^{LM}_n&=&(-1)^n4^n(L^{LM})^{n+1/2}_{>0},
\eea
where $\mu=\mu(x,\th)$ and $\tau=\tau(x,\th)$ are even and odd superfields, respectively.
It should be mentioned that the LM theory discussed here is obtained from the
$N=2, \alpha=-2$ LM sKdV theory \cite{LM}. The bi-Hamiltonian structure of the LM 
hierarchy is also taken from \cite{OP}, in component form, as \cite{MP1}
\bea
(\Th^{LM}_1)^{-1}& : &
\left(
\ba{c}
\dot u\\
\dot \tau
\ea
\right)
\rightarrow
\left(
\ba{c}
\de u\\
\de \tau
\ea
\right)=
\left(
\ba{cc}
D-D^{-1}\tau D^{-1} & u\pa^{-1}+D^{-1}uD^{-1}\\
\pa^{-1}u+D^{-1}uD^{-1} & D^{-1}-\pa^{-1}\tau\pa^{-1}
\ea
\right)
\left(
\ba{c}
\dot u\\
\dot \tau
\ea
\right)
\label{hlam1}\\
\Th^{LM}_2& : &
\left(
\ba{c}
\de u\\
\de \tau
\ea
\right)
\rightarrow
\left(
\ba{c}
\dot{u}\\
\dot{\tau}
\ea
\right)=
\left(
\ba{cc}
-D\pa+\tau & 2u\pa-(Du)D+2u_x\\
 2u\pa-(Du)D+u_x & -D\pa^2+3\tau\pa+(D\tau)D+2\tau_x
\ea
\right)
\left(
\ba{c}
\de u\\
\de \tau
\ea
\right)
\label{hlm2}
\eea
where, similarly, the phase space of the LM theory can be represented as a set of
pairs $n=(u,\tau)$. Then the tangent and cotangent vectors at a point $n$ are 
represented as $\dot{n}=(\dot{u},\dot{\tau})$ and $\de n=(\de u,\de\tau)$, 
respectively, where $\dot{u}$ and $\de\tau$ are even, while $\de u$ and 
$\dot{\tau}$ are odd. The inner product is defined by
$\langle \de n, \dot{n}\rangle\equiv \int dx d\th (\de u\dot{u}+\de \tau\dot{\tau})$.
More features about the bi-Hamiltonian structures of these two hierarchies
have been tabulated in Ref.\cite{MP1}.

\section{Gauge transformations}

In Ref.\cite{IK}, Inami and Kanno showed that the MR sKdV hierarchy 
can be related to the LM sKdV hierarchy via the following gauge transformation:
\be
L^{MR}_1=S_1^{-1}L^{LM}S_1,\qquad S_1=e^{\int^x u},
\label{s1}
\ee
which leads to
\be
\phi_1=(Du)+\tau,\qquad a_1=u_x-u^2-((Du)+\tau)(D^{-1}u).
\label{mur1}
\ee
They also showed that the Lax equation (\ref{eqmr}) of the LM theory 
is mapped into the Lax equation (\ref{eqlm}) of the MR theory under such transformation.
Hence (\ref{mur1}) provides a gauge equivalence between these two hierarchies and now 
is referred to as the Inami-Kanno transformation. It can be shown that $S_1^{-1}$ is an 
eigenfunction of the MR sKdV hierarchy, i.e. $\pa_nS_1^{-1}=(B^{MR}_nS_1^{-1})$.
Furthermore, Morosi and Pizzocchero \cite{MP1}
showed that the IK transformation is a canonical map, in the sense that the bi-Hamiltonian
structure of the LM sKdV hierarchy is mapped to the bi-Hamiltonian structure of
the MR sKdV hierarchy. That is,
\bea
 S_1'^{\dag}(\Th^{MR}_1)^{-1}S_1' &=& (\Th^{LM}_1)^{-1} 
\label{can1}\\
S_1'(\Th^{LM}_2)S_1'^{\dag}&=& (\Th^{MR}_2), 
\label{can2}
\eea
where $S_1'$ and $S_1'^{\dag}$ are linearized map and its transport map respectively 
of the IK transformation and satisfy  
\be
\langle S_1'^{\dag}\de m, \dot{n}\rangle=\langle \de m, S'_1\dot{n}\rangle.
\ee

In fact, we can construct another transformation between MR and LM sKdV hierarchies
as follows:
\be
L^{MR}_2=S_2^{-1}L^{LM}S_2,\qquad S_2=D^{-1}S_1.
\label{s2}
\ee
Then a simple calculation leads to
\be
\phi_2=(Du)-\tau,\qquad a_2=-u^2-(D\tau)-((Du)-\tau)(D^{-1}u).
\label{mur2}
\ee
It can be shown that the Lax formulations are preserved under such 
transformation. Hence the transformation (\ref{s2}) also provides a
gauge equivalence of the MR and LM sKdV hierarchies. Similarly, we
can show that, in this case,  $\pa_nS_1=-((B^{MR}_n)^*S_1)$.
That means $S_1$ is an adjoint eigenfunction of the
MR sKdV hierarchy. 

Next, let us discuss the canonical property of the transformation (\ref{s2}). 
From (\ref{mur2}), the linearized map $S_2'$ and its adjoint map $S_2'^{\dag}$ 
can be derived as follows:
\bea
S_2' &=&
\left(
\ba{cc}
-2u+(D^{-1}u)D-\phi_2 D^{-1} & -D-(D^{-1}u)\\
D & -1
\ea
\right)
\label{lin}\\
S_2'^{\dag}&=&
\left(
\ba{cc}
-2u+D^{-1}\phi_2-D( D^{-1}) & -D\\
-D+(D^{-1}u) & -1
\ea
\right).
\label{alin}
\eea
After a straightforward but tedious calculation, 
the Poisson structures transform as
\bea
 S_2'^{\dag}(\Th^{MR}_1)^{-1}S_2' &=&-(\Th^{LM}_1)^{-1} \\
S_2'(\Th^{LM}_2)S_2'^{\dag}&=& -(\Th^{MR}_2),
\eea
which, comparing with (\ref{can1}) and (\ref{can2}), acquire a minus sign. 
It seems that this result contradicts the property of preserving
the Lax formulations. However, it is not the case. Since the parity
of the gauge operator $S_2$ is odd, the Hamiltonian
of the LM hierarchy $H_n^{LM}=\str((L^{LM})^{n+1/2})$ 
(up to a multiplicative constant) then is transformed to
$H_n^{MR}=-H_n^{LM}$ due to the following property:
\be
\str(PQ)=(-1)^{|P||Q|}\str(QP).
\ee
where $P$ and $Q$ are any super-pseudo-differential operators with gradings
$|P|$ and $|Q|$, respectively.
Therefore, the gauge equivalence is compatible with the
canonical property under the transformation triggered by the gauge operator $S_2$. 

Based on the above discussions, the canonical property of the gauge transformations
between the MR and LM sKdV hierarchies can be summarized as follows,
\bea
S_i'^{\dag}(\Th^{MR}_1)^{-1}S_i' &=& (-)^{|S_i|}(\Th^{LM}_1)^{-1} \\
S_i'(\Th^{LM}_2)S_i'^{\dag}&=& (-1)^{|S_i|}(\Th^{MR}_2), \qquad i=1,2
\eea
which seems to be the supersymmetric generalization of the 
bosonic case \cite{ST}.

\section{B\"acklund transformation and Kupershmidt-Wilson theorem}

From the IK transformation, we know that if we have a solution $\{u,\tau\}$ of the LM sKdV
hierarchy, then Eq. (\ref{mur1}) gives a solution $\{\phi_1,a_1\}$ of the MR sKdV
hierarchy. Sometimes, such a transformation of one hierarchy to another is called
a Miura transformation. On the other hand, Eq.(\ref{mur2}) also gives another
solution $\{\phi_2,a_2\}$ of the MR sKdV hierarchy. Hence a Darboux-B\"acklund
transformation (DBT) of the MR sKdV hierarchy to itself can be constructed
from these two gauge transformations.
In other words, let $\{\phi_1,a_1\}$ be a solution of the MR sKdV
hierarchy, then solving $\{u,\tau\}$ from (\ref{mur1}) and substituting it into (\ref{mur2})
we get
\bea
\phi_2 &=& -\phi_1-2(D^3\ln S_1^{-1})\\
a_2 &=& a_1-(D\phi_1)+2(D\ln S_1^{-1})(\phi_1+(D^3\ln S_1^{-1})),
\eea
which is just the DBT derived in Ref.\cite{Liu}. The action of the gauge operators 
$S_1$ and $S_2$ for the MR and LM sKdV hierarchies are shown as follows

\begin{center}
\be
\ba{ccccc}
  & & LM & & \\
 S_1& \swarrow& & \searrow&S_2 \\
 MR^1&\longleftarrow & DBT &\longrightarrow &MR^2 
\ea
\label{relation}
\ee
\end{center}

In the following, we want to discuss the relationship in (\ref{relation}) 
from the KW theorem viewpoint, in which the gauge operator $S_2$ 
plays an important and unambiguous role.
From (\ref{laxlm}), the Lax operator $L^{LM}$ can be factorized as follows, \cite{IK}
\bea
L^{LM}&=&\pa^2-2u\pa-((Du)+\tau)D\no\\
&=&(D-\Phi_1)(D-\Phi_1-\Phi_2)(D-\Phi_2)D 
\label{faclm}
\eea
where $u$ and $\tau$ can be expressed in terms of  the superfields $\Phi_i$ as 
\bea
u &=&\frac{1}{2}[(D\Phi_1)+(D\Phi_2)-\Phi_1\Phi_2]
\label{for1}\\
\tau &=&\frac{1}{2}[\Phi_{2x}-\Phi_{1x}-(D\Phi_1)\Phi_2-\Phi_1(D\Phi_2)]
\label{for2}
\eea
The second Hamiltonian structure of the LM theory can be simplified under the 
factorization (\ref{faclm}). 
From (\ref{for1}) and (\ref{for2}), it is straightforward to show that
\be
\Th^{LM}_2=[\frac{\pa(u,\tau)}{\pa(\Phi_1,\Phi_2)}]
\left(
\ba{cc}
0 & 2D\\
2D & 0
\ea
\right)
[\frac{\pa(u,\tau)}{\pa(\Phi_1,\Phi_2)}]^{\dagger}
\ee
where the Fr\'echet derivative can be calculated as:
\be
[\frac{\pa(u,\tau)}{\pa(\Phi_1,\Phi_2)}]
=
\left(
\ba{cc}
-\frac{1}{2}(D+\Phi_2) & -\frac{1}{2}(D-\Phi_1) \\
-\frac{1}{2}(\pa+\Phi_2D+(D\Phi_2)) & \frac{1}{2}(\pa-\Phi_1D-(D\Phi_1)) 
\ea
\right)\\
\ee
and $[\frac{\pa(u,\tau)}{\pa(\Phi_1,\Phi_2)}]^{\dagger}$ is its formal adjoint.

Now applying the IK transformation to (\ref{faclm}), we obtain the multiplicative form
of the Lax operator $L_1^{MR}$ as
\be
L_1^{MR}=(D-\Psi_1)(D-\Psi_2 )(D-\Psi_3 )(D-\Psi_4), 
\ee
where the superfields $\Psi_i$ are given by
\bea
\Psi_1&=&\frac{1}{2}((D^{-1}\Phi_1\Phi_2)+\Phi_1-\Phi_2)\\
\Psi_2&=&\frac{1}{2}((D^{-1}\Phi_1\Phi_2)+\Phi_1+\Phi_2)\\
\Psi_3&=&\frac{1}{2}((D^{-1}\Phi_1\Phi_2)+\Phi_2-\Phi_1)\\
\Psi_4&=&\frac{1}{2}((D^{-1}\Phi_1\Phi_2)-\Phi_1-\Phi_2).
\eea
where only two of them are independent variables. The Lax equation for
$L_1^{MR}$ then can be expressed in terms of the hierarchy equations
of $\Psi_i$.

On the other hand, if we apply the gauge transformation (\ref{s2}) to (\ref{faclm}), 
the Lax operator $L_2^{MR}$ then is factorized as 
\be
L_2^{MR}=(D-\Psi_4)(D-\Psi_1 )(D-\Psi_2 )(D-\Psi_3), 
\ee
which differs from $L_1^{MR}$ only by a cyclic permutation: 
$\Psi_1\mapsto \Psi_2, \cdots, \Psi_4\mapsto\Psi_1$. 
Such cyclic permutation does not change the hierarchy equations of $\Psi_i$ \cite{A}
and hence  generates the DBT for the MR sKdV hierarchy itself.

\section{Concluding remarks}

We have shown that, in addition to the IK transformation, there is another
gauge transformation between the MR and LM sKdV hierarchies. We investigate 
the canonical property of this new gauge transformation and show that
it depends on the grading (or parity) of the gauge operator. 
Using this new gauge transformation and the IK transformation we rederived the DBT
for the MR sKdV hierarchy. We also give an interpretation of this new 
gauge transformation from the KW theorem viewpoint. 

Finally, we would like to remark that the geometrical interpretation of the
IK transformation discussed in Ref.\cite{MP1} can be applied to the new
gauge transformation as well. The only thing we have to do is to choose
a different cross section $\hat{\Sigma}$, which is matrix in the 
fiber over $m$ of the form
\be
\hat{\Sigma}(m)=
\left(
\ba{cccc}
0  &0 &1 & 0\\
(Du)-\tau &0 & 0& 1\\
-2u &  -1 & 0 &0\\
0& 0& 0& 0
\ea
\right).
\ee
Then the transformation (\ref{mur2}) comes out naturally from a general 
equation derived in Ref.\cite{MP1} which describs the quotient space in the
bi-Hamiltonian reduction of a loop superalgebra. Since the IK transformation 
was also derived from the same equation, thus (\ref{mur1}) and (\ref{mur2}) can 
be treated on an equal footing in the bi-Hamiltonian framework.

{\bf Acknowledgments\/}
This work is supported by the National Science Council of Taiwan 
under grant No. NSC-87-2811-M-007-0025.
M.H. Tu also wish to thank Center for Theoretical Sciences of National
Science Council of Taiwan for partial support.

\end{document}